# Large Anomalous Hall Effect in a Noncoplanar Magnetic Heterostructure

Anke Song,[1] Jine Zhang,[2] Yequan Chen,[1] Zhizhong Zhang,[2] Xinjuan Cheng,[3] Ruijie Xu,[1] Wenzhuo Zhuang,[1] Wenxuan Sun,[1] Yong Zhang,[1] Xu Zhang,[1] Zhongqiang Chen,[1] Fengqi Song,[4,5] Yue Zhang,[2,*] Xuechao Zhai,[3,*] Yongbing Xu,[1] Weisheng Zhao,[2] Rong Zhang,[1,6] and Xuefeng Wang[1,5,*]

[1]Jiangsu Provincial Key Laboratory of Advanced Photonic and Electronic Materials, State Key Laboratory of Spintronics Devices and Technologies, School of Electronic Science and Engineering, Collaborative Innovation Center of Advanced Microstructures, Nanjing University, Nanjing 210093, China

[2]School of Integrated Circuit Science and Engineering, Beihang University, Beijing 100191, China

[3]MIIT Key Laboratory of Semiconductor Microstructures and Quantum Sensing, School of Physics, Nanjing University of Science and Technology, Nanjing 210094, China

[4]National Laboratory of Solid State Microstructures, School of Physics, Nanjing University, Nanjing 210093, China

[5]Atom Manufacturing Institute, Nanjing 211806, China

[6]Department of Physics, Xiamen University, Xiamen 361005, China

[*]Authors to whom correspondence should be addressed. E-mail: yz@buaa.edu.cn (Y. Zhang); zhaixuechao@njust.edu.cn (X. Zhai); xfwang@nju.edu.cn (X. Wang)

**Abstract**

The anomalous Hall effect (AHE) occurs in magnetic systems and also unexpectedly in non-magnetic materials adjacent to magnetic insulators via the heterointerface interactions. However, the AHE in heterostructures induced by magnetic proximity effect remains quite weak, restricting their practical device applications. Here, we report a large intrinsic AHE with a resistivity of 114 n$\Omega$ cm at 5 K in noncoplanar magnetic heterostructures of $Cr_5Te_6$/Pt. This is the record-high AHE value among all the magnetic insulators/heavy metal heterostructures. A reversal of the AHE signal occurs due to the reconstruction of Berry curvature at the Fermi level, which is verified by the first-principles calculations. Topological spin textures at the interface are directly visualized via high-magnetic-field magnetic force microscopy, which accounts for the large AHE, as confirmed by the atomic simulations. These findings open a new avenue for exploring the large AHE in heterointerfaces and facilitate the potential applications in topological spintronic devices.

## 1. Introduction

The anomalous Hall effect (AHE) is one of the fundamental and widely investigated phenomena in physics and electronics. Due to the time-reversal symmetry breaking and spin-orbit interaction, AHE is typically dominated by ferromagnetism and is generally proportional to the magnetization.[1] The origin of AHE in ferromagnet is broadly ascribed to the intrinsic mechanism associated with Berry curvature[2] or the extrinsic mechanism arising from asymmetric scattering caused by structural defects and magnetic impurities.[1] The intrinsic AHE is governed by the electronic structures of the material, where the Berry curvature occupying the Bloch bands corresponds to an effective magnetic field in momentum space, allowing electrons to acquire transverse momentum during motion along the direction of electric field.[3-4] This mechanism is also applicable to describe AHE in other systems rather than ferromagnetic materials. Recently, sizable AHE has been observed in systems with nearly zero magnetization, such as spin liquids,[5] antiferromagnetic systems (both non-collinear[6-8] and collinear[9-10]), and even non-magnetic topological material $ZrTe_5$,[11] driven by the non-zero Berry curvature in momentum space.

Since the discovery of magnetic skyrmions, topological spin textures have persisted at the forefront of spintronics research.[12-13] These vortex-like quasiparticles can carry topological charge and show substantial potential in advanced data storage and neuromorphic computing applications.[14] In recent years, the role of topological spin textures in the AHE has attracted substantial attention. On the one hand, spin textures with scalar spin chirality in frustrated magnets[15-17] can act as skew scattering centers, causing carriers to acquire transverse momentum. Additionally, in certain antiferromagnets,[6-7, 18-20] the interaction between spin textures and the electronic band structures can induce non-zero Berry curvature, leading to the intrinsic AHE. Nevertheless, investigations on the contribution of such spin textures to AHE primarily focus on the individual material systems, while AHE induced by topological spin textures at heterointerfaces remains unexplored. Heterostructures composed of magnetic insulators/heavy metal (MI/HM) with strong spin-orbit coupling (SOC) have been widely investigated.[13, 21] In these systems, AHE arising from magnetic proximity effects can be realized. Generally, a large AHE is favorable for low-power applications since the Joule heating of the longitudinal current can be effectively suppressed.[22] However, the AHE in MI/HM systems remains quite weak, often around 10 nΩ cm,[21, 23] which seriously limits the practical applications.

The low-dimensional magnetic $Cr_xTe_y$ system features the alternating Cr-deficient and Cr-full layers stacking along the *c*-axis direction, possessing rich magnetic properties and electronic band structures,[22, 24-27] including topological spin textures[25] and tunable Berry curvature.[24, 27] This makes $Cr_xTe_y$ an ideal material platform for exploring the AHE. In this work, we observe a large intrinsic AHE in the noncoplanar magnetic $Cr_5Te_6$ heterostructures enabled by topological spin textures. The extracted anomalous Hall resistivity reaches 114 nΩ cm at 5 K, which is the record-high value among the MI/HM heterostructures. The reversal of the AHE is ascribed to the reconstruction of Berry curvature, which is verified by our first-principles calculations. Topological spin textures at the heterointerface are directly visualized from the low-temperature (1.6 K), high-magnetic-field (9 T) magnetic force microscopy (MFM). The atomic simulations further demonstrate that the interfacial Dzyaloshinskii-Moriya interaction (DMI) stabilizes the topological spin textures, driven by the SOC and broken inversion symmetry. The created interfacial topological spin textures are anticipated to extend to other heterostructures for chiral spintronics.

## 2. Results and Discussion

Epitaxial $Cr_5Te_6$ films are fabricated on cleaned $Al_2O_3$ (0001) substrates by the pulsed laser deposition (PLD) technique. The crystal structure of $Cr_5Te_6$ exhibits a monoclinic phase resembling that of $CrTe_2$ with a space group of *I2/m-C$_{2h}$*.[28] The schematic diagram of the heterostructure is shown in **Figure 1**a. $Cr_5Te_6$ is a self-intercalated phases of $Cr_xTe_y$, consisting of the host two-dimensional $CrTe_2$ layers with Cr atoms intercalated in the van der Waals gaps. The x-ray diffraction (XRD) pattern (Figure 1b) demonstrates the single-crystal nature along the *c*-axis orientation of the $Cr_5Te_6$ films. The atomic force microscope (AFM) image (inset of the Figure 1b) reveals a relatively flat surface morphology of the $Cr_5Te_6$ films (the root mean square roughness ~0.4 nm). Moreover, the cross-sectional high-resolution scanning transmission electron microscopy (STEM) image (Figure 1c) shows a thin layer of amorphous Pt (~3 nm) deposited on the 10-nm-thick $Cr_5Te_6$ film surface. The corresponding energy dispersive x-ray spectroscopy (EDX) elemental mapping in the right panel indicates a sharp interface without the obvious elemental interdiffusion.

The magnetic properties of the 10-nm-thick $Cr_5Te_6$ films are measured in both in-plane (IP) and out-of-plane (OOP) configurations using a superconducting quantum interference device (SQUID). The temperature-dependent magnetization curves of IP and OOP (see Figure S1a, b, Supporting Information) exhibit abnormal kinks, indicating a temperature-dependent interaction between antiferromagnetic and ferromagnetic phases.[28-29] This complex interaction is attributed to the critical balance between Pauli repulsion and kinetic energy gain across the layers.[30-31] In addition, previous studies on $Cr_xTe_y$ manifest that both the Néel temperature and the Curie temperature increase with increasing Cr content.[25, 32] Consistently, our $Cr_5Te_6$ possesses a relatively high Curie temperature (320 K).[28] The *M-H* curves show the coercive fields of about 300 Oe at 5 K (see Figure S1c, d, Supporting Information), reflecting the inherent noncoplanar magnetic structure of $Cr_5Te_6$.[25, 28] The *M-H* curves in different directions at 5 K and 300 K (see Figure S1e, Supporting Information) illustrate the approximate isotropy of the 10-nm-thick $Cr_5Te_6$ films. Notably, a previous report showed that the antiferromagnetic phase increases as the film thickness decreases in the $Cr_5Te_8$ system.[33] Likewise, a similar dimensional effect also exists in the 10-nm-thick $Cr_5Te_6$ films, leading to an obvious noncoplanar magnetic structure compared to the thicker samples.

To investigate the potential AHE in the $Cr_5Te_6$-based heterostructures, 10-nm-thick $Cr_5Te_6$ films are chosen as the magnetic layers owing to its extremely high resistance (see Figure S2, Supporting Information), excluding the parasitic conduction in the $Cr_5Te_6$ films. The $Cr_5Te_6$ (10 nm)/Pt heterostructures with different Pt thickness are fabricated into the standard Hall-bar devices (Figure 1d). The temperature-dependent longitudinal resistivities ($\rho_{xx}$) of the various $Cr_5Te_6$/Pt devices display typical metallic behavior (Figure 1e). However, an anomalous upward trend is observed around 30 K due to the scattering effects at the interface.[34] Notably, this scattering effect diminishes as the thickness of Pt ($t_{Pt}$) increases. The $\rho_{xx}$ as a function of thickness (Figure 1f) is fitted by the semiclassical theoretical model according to Fuchs-Sondheimer theory:[35]

$$\rho_{Pt} = \rho_\infty \left[1 - \left(\frac{1}{2} + \frac{3}{4}\frac{\lambda}{t_{Pt}}\right)\left(1 - pe^{-\frac{\zeta t_{Pt}}{\lambda}}\right)e^{-\frac{t_{Pt}}{\lambda}}\right]^{-1} \quad (1)$$

where $\rho_\infty$ is the bulk resistivity of Pt, $\lambda$ is the mean free path, $p$ is the scattering coefficient of electrons by the film, and $\zeta$ is the scattering coefficient of electrons by the grain boundaries. The well-fitted curve with the parameters of $\rho_\infty$ = 34 μΩ cm, $\lambda$ = 8.4 nm, $p$ = 0.75, and $\zeta$ = 0.25[36] demonstrates that the current flows only through the uniform Pt layer.

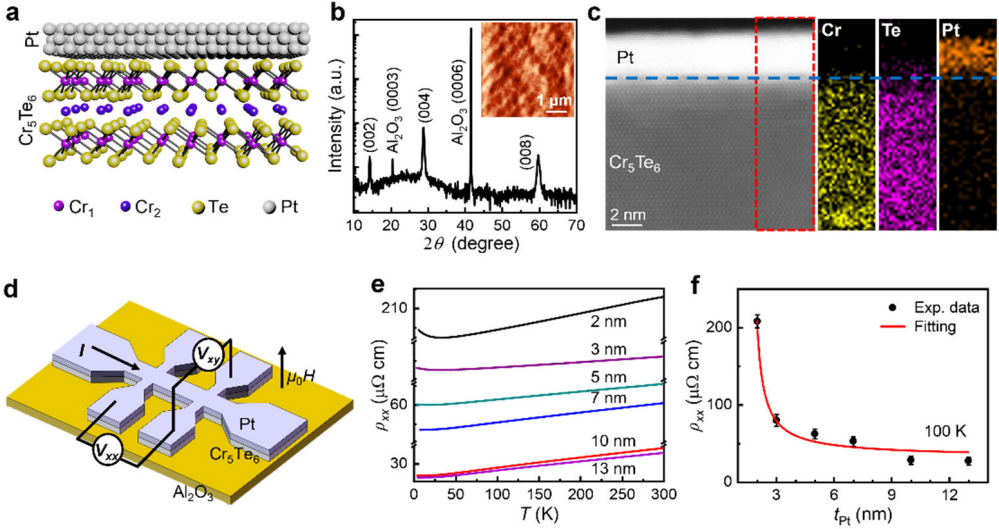

**Figure 1.** Basic structural and longitudinal transport properties of $Cr_5Te_6$/Pt heterostructures. a) Schematic diagram of the $Cr_5Te_6$/Pt. $Cr_5Te_6$ is regarded as a monoclinic $CrTe_2$ structure with an additional Cr intercalation. b) The typical XRD pattern of $Cr_5Te_6$ with the thickness of 20 nm. The inset shows an AFM image of the surface of $Cr_5Te_6$ with root mean square of 0.4 nm. c) The cross-sectional STEM image of the $Cr_5Te_6$ (10 nm)/Pt (3 nm) heterostructure taken along [110] direction. The right panel is the corresponding EDX elemental mapping of Cr, Te, and Pt at the interface marked by the red boxed area. d) Schematic configuration of a Hall bar device. e) Temperature dependence of longitudinal resistivity ($\rho_{xx}$) for $Cr_5Te_6$ (10 nm)/Pt devices with the Pt thickness ranging from 2 to 13 nm. f) The Pt thickness dependence of the $\rho_{xx}$ at 100K.

Hall measurements are implemented on various devices. **Figure 2**a presents the Hall curves of the Cr$_5$Te$_6$ (10 nm)/Pt (3 nm) heterostructure (termed as Device 1) at varying temperatures. In general, the Hall resistivity ($\rho_{xy}$) arises from the contributions of the ordinary Hall effect (OHE) and the AHE, as follows:[19]

$$\rho_{xy}(H) = \rho_{xy}^{\text{OHE}} + \rho_{xy}^{\text{AHE}} \qquad (2)$$

The distinct AHE is observed from 5 to 300 K, with the anomalous Hall resistivity ($\rho_{xy}^{\text{AHE}}$) decreasing monotonically with increasing temperature. As a non-magnetic material, the bare Pt does not exhibit the AHE hysteresis loop. Therefore, the emergence of AHE in Pt should arise from the interfacial effect. At 5 K, $\rho_{xy}^{\text{AHE}}$ reaches a maximum of 114 nΩ cm. A marked decline in $\rho_{xy}^{\text{AHE}}$ occurs between 200 and 300 K with a reversal of the AHE signal (see Figure S3, Supporting Information). Figure 2b illustrates the temperature-dependent Hall curves in the Cr$_5$Te$_6$/Pt (10 nm) heterostructure, indicating a weaker AHE signal as compared to Device 1. Similar phenomenon is also observed in the Cr$_5$Te$_6$/Pt (5 nm) and Cr$_5$Te$_6$/Pt (7 nm) heterostructures (see Figure S4a, b, Supporting Information). When the thickness of Pt increases up to 13 nm, the Hall signal only exhibits an OHE (see Figure S4c, Supporting Information). It is worth noting that the Cr$_5$Te$_6$/Pt (2 nm) exhibits a similar OHE. This might be due to the three-dimensional island-like growth mode of Cr$_5$Te$_6$, which makes it impossible to form a completely uniform film below the 2-nm-thick Pt layer. This is unfavorable for the formation of spin textures at the interface (see Figure S5, Supporting Information). To confirm that the AHE is dominated by the interfacial effect, a layer with the weak SOC (e.g., Cu) is inserted at the interface to isolate the electron wave functions.[37] As illustrated in Figure 2c, a linear Hall signal is observed in the Cr$_5$Te$_6$/Cu/Pt heterostructure, unambiguously demonstrating that the AHE originates from the Cr$_5$Te$_6$/Pt interface. The comparison of the AHE in the Cr$_5$Te$_6$/Pt (3 and 10 nm) heterostructures at 5 K reveals the intriguing reversal of the AHE signal (Figure 2d). Similar phenomenon also occurs in Cr$_5$Te$_6$/Pt (5 and 7 nm) (see Figure S4d, Supporting Information) as will be elaborated on later.

When the Pt layer exceeds 3 nm, a distinct hump appears in the AHE curve, signifying the existence of both electron and hole carriers. Previously, the angle resolved photoemission spectroscopy on Cr$_{1+\delta}$Te$_2$[24] and the electrical transport of Cr$_5$Te$_6$[28] substantiated the existence of substantial holes in Cr$_5$Te$_6$ which arise from the unoccupied *d*-orbitals of Cr atoms.[38] As a result, the holes completely dominate the transport behavior in Cr$_5$Te$_6$/Pt (3 nm) heterostructure. However, in the thicker Pt (5-10 nm), both holes and electrons contribute to the transport and induce the hump in the Hall curve. The nonlinear ordinary Hall effect (NLOHE) curve is attributed to the resultant Lorentz deflections between holes and electrons. The Hall curves (Figure 2e) consist of contributions from the NLOHE and AHE, which can be fitted by the following equation:[39-41]

$$\rho_{xy}(H) = \rho_{xy}^{\text{NLOHE}} + \rho_{xy}^{\text{AHE}}$$
$$= \frac{\mu_0 H}{e} \frac{(p\mu_h^2 - n\mu_e^2) + (\mu_0 H)^2 \mu_h^2 \mu_e^2 (p-n)}{(p\mu_h - n\mu_e)^2 + (\mu_0 H)^2 \mu_h^2 \mu_e^2 (p-n)^2} + M_0 \tanh\left(\frac{\mu_0 H}{a_0} - \mu_0 H_0\right) \qquad (3)$$

where $e$ is the fundamental electron charge, $\mu_h$ ($\mu_e$) and $p$ ($n$) are the mobility and carrier density of holes (electrons), respectively, while $M_0 \tanh\left(\frac{\mu_0 H}{a_0} - \mu_0 H_0\right)$ is the AHE contribution. The superposition of these two Hall effects is also observed in Dirac material $La_3MgBi_5$.[42] The carrier densities of electrons and holes in the $Cr_5Te_6$/Pt (10 nm) heterostructure are estimated to be $4.50 \times 10^{22}$ cm$^{-3}$ and $1.74 \times 10^{21}$ cm$^{-3}$, respectively. Similar electron-hole interdiffusion and signal reversal of the AHE are found in the $Sb_2Te_3$/$Sb_{1.9}V_{0.1}Te_3$[43] and $ZrTe_2$/$CrTe_2$[44] systems. The hump is gradually suppressed and completely vanishes at 200 K (Figure 2b) due to the upward shift of the Fermi level with increasing temperature, resulting in a transport regime that is entirely electron-dominated. To investigate the mechanism of the AHE, it is necessary to reveal the scaling relation between the $\rho_{xy}^{AHE}$ and the $\rho_{xx}$.[43] The quadratic dependence indicates the intrinsic[1] or extrinsic side-jump mechanisms,[45] while the linear relationship corresponds to the extrinsic skew scattering mechanism.[1] In Figure 2f, the linear relation between $\rho_{xy}^{AHE}$ and $\rho_{xx}^2$ corroborates the prominent influence of intrinsic or side-jump mechanism. Here, we emphasize that the intrinsic mechanism dominates the AHE of the $Cr_5Te_6$/Pt heterostructures due to the strong SOC in Pt.[46] Thus, the electron-hole interdiffusion and the varying temperature both manipulate the Fermi level, leading to the reconstruction of Berry curvature and the large AHE.[43] This topological amplification of AHE could be further enhanced through hole doping in the correlated oxide system ($La_{0.7}Sr_{0.3}MnO_3$/$SrIrO_3$).[47]

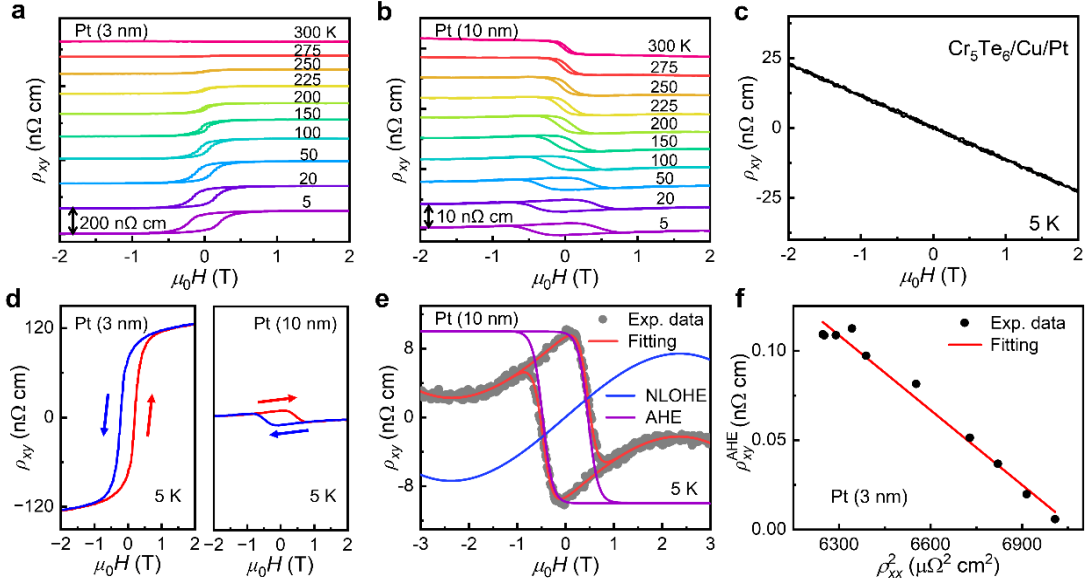

**Figure 2.** The AHE observed in $Cr_5Te_6$ (10 nm)/Pt heterostructures. a, b) The field-dependent Hall resistivity at various temperatures in $Cr_5Te_6$/Pt (3 nm) and $Cr_5Te_6$/Pt (10 nm) heterostructures, respectively. c) The field-dependent Hall resistivity of $Cr_5Te_6$/Pt (3 nm) with the interfacial insertion of Cu (5 nm) at 5 K. d) The observation of the AHE signal reversal in the $Cr_5Te_6$/Pt heterostructures at 5 K. e) The fitting of the Hall signal in $Cr_5Te_6$/Pt (10 nm) heterostructure, yielding the separated $\rho_{xy}^{AHE}$ and $\rho_{xy}^{NLOHE}$. f) The plot of the $\rho_{xy}^{AHE}$ as a function of the $\rho_{xx}$ squared.

The temperature-dependent $\rho_{xy}^{AHE}$ of various MI/HM heterostructures is summarized in **Figure 3**, including Y$_3$Fe$_5$O$_{12}$ (YIG)/W,[48] YIG/Pt[49], Tm$_3$Fe$_5$O$_{12}$ (TmIG)/W,[50] TmIG/Pt,[21] Y$_{3-x}$Tm$_x$Fe$_5$O$_{12}$ (YTmIG)/Pt,[51] Cr$_2$O$_3$/Pt,[46] NiO/Pt,[52] Cr$_2$Ge$_2$Te$_6$ (CGT)/Pt,[53] and CGT/W.[54] The Device 1 in our work attains the maximum 114 nΩ cm at 5 K, maintaining its great prominence among all the MI/HM heterostructures. Moreover, methods such as ionic-gating technique are expected to further enhance and regulate the AHE of the Cr$_5$Te$_6$-based heterostructure.

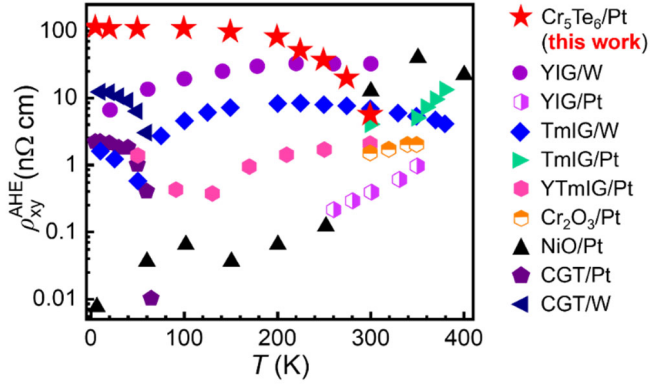

**Figure 3.** Summary of the anomalous Hall resistivity in the different heterostructures. The temperature dependences of the observed saturated $\rho_{xy}^{AHE}$ values in Cr$_5$Te$_6$/Pt (this work) and other MI/HM heterostructures, such as YIG/W,[48] YIG/Pt[49], TmIG/W,[50] TmIG/Pt,[21] YTmIG/Pt,[51] Cr$_2$O$_3$/Pt,[46] NiO/Pt,[52] CGT/Pt,[53] and CGT/W.[54]

To reveal the intrinsic mechanism of AHE from the perspective of symmetry, we perform the first-principles calculations of electronic properties based on a minimal Cr$_5$Te$_6$/monolayer Pt structure (**Figure 4**a). The high-symmetry momentum direction X–Γ–Y and its time-reversal (or spatial inversion) path X'–Γ–Y' are utilized to present the electronic band structures and Berry curvature (Figure 4b). Since time-reversal symmetry necessarily requires $\Omega_n(\boldsymbol{k})=-\Omega_n(-\boldsymbol{k})$ and inversion symmetry definitely requires $\Omega_n(\boldsymbol{k})=\Omega_n(-\boldsymbol{k})$ to ensure the invariance of group velocity within topological theory,[1] the result of $\Omega_n(\boldsymbol{k})\neq -\Omega_n(-\boldsymbol{k})$ and $\Omega_n(\boldsymbol{k})\neq\Omega_n(-\boldsymbol{k})$ here undoubtedly demonstrates the simultaneous breaking of time-reversal symmetry and inversion symmetry. It is clear that the non-magnetic Pt layer with strong SOC has time-reversal symmetry, which nevertheless is broken by the adjacent magnetic Cr$_5$Te$_6$ layer. By integrating $\Omega_n(\boldsymbol{k})$ over the first Brillouin zone for all the occupied bands, the anomalous Hall conductivity $\sigma_{AHE}$ is derived in Figure 4c. The individual Pt layer exhibits no AHE due to its time-reversal symmetry. Despite the symmetry is broken in Cr$_5$Te$_6$ layer, it displays a weaker AHE on the order of $10^0$-$10^2$ Ω$^{-1}$ cm$^{-1}$ near the Fermi energy. However, when coupled with the Pt layer, the AHE is significantly enhanced by an order of magnitude below the Fermi energy (1-2 meV). Thus, the conductive Pt layer dominates the AHE here. Cr$_5$Te$_6$ is particularly effective for generating large Berry curvature peaks (Figure 4c) when in contact with Pt, as it provides significant Cr atoms per unit area—2 to 3 times larger than those in Cr$_2$O$_3$/Pt[46] and CGT/Pt.[53] This is further evidenced by a substantial charge accumulation of 0.15 electrons per Pt atom on average in monolayer Pt (see the details in the inset of Fig. 4c), which decreases to 0.09 electrons per Pt atom when bilayer Pt is in contact with $Cr_5Te_6$ (see Figure S6, Supporting Information). As the thickness of Pt changes from monolayer Pt (3.04 Å) to bilayer Pt (5.47 Å), the AHE behavior only undergoes a slight change in amplitude, which is due to the decrease in interlayer charge accumulation. According to the existing theoretical calculations,[53] the extent of induced magnetization penetration into the Pt layer is about 1.2 nm (four Pt layers).

Notably, the reversal of the AHE as the temperature increases (see Figure S3, Supporting Information) is attributed to the reconstruction of Berry curvature. The positive linear Hall curves under high magnetic fields show that only holes participate in the transport (Figure S3a-c). However, when the temperature rises to 300 K, the Hall curves under high magnetic fields exhibit the nonlinear behavior (Figure S3d), which is caused by the simultaneous participation of electrons and holes (similar to Figure 2e). Thus, the Fermi level of the heterostructure gradually shifts upward from the valence band to the conduction band as the temperature increases. Figure 4 theoretically displays that the anomalous Hall conductivity induced by the Berry curvature undergoes a sign reversal and experiences a significant decrease in magnitude with the upward shift of the Fermi level.

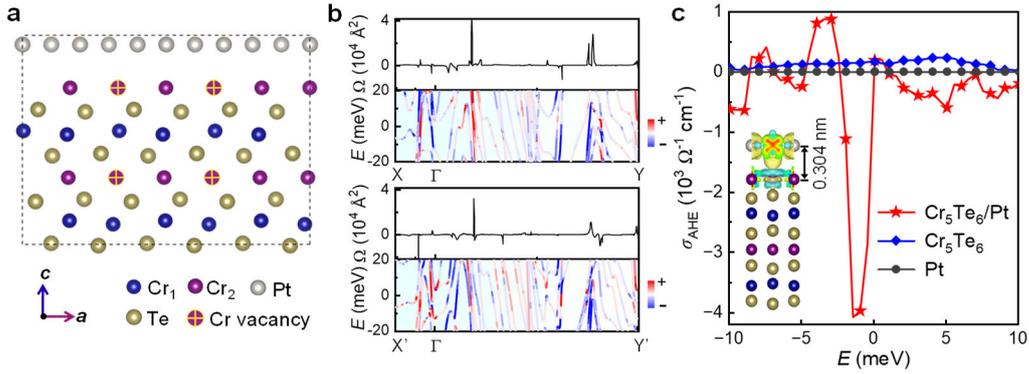

**Figure 4.** The theoretical calculations of Berry curvature and anomalous Hall conductivity in $Cr_5Te_6$/Pt. a) The calculated interface structure of $Cr_5Te_6$/Pt (monolayer). b) Berry curvature and the corresponding band structures along the high-symmetry momentum direction X–Γ–Y and its time-reversal path X'–Γ–Y'. The color label on the right represents the positive and negative Berry curvature. c) Dependence of the anomalous Hall conductivity on the Fermi level. The results for individual $Cr_5Te_6$ and Pt are plotted here for comparison. The inset shows the charge density difference, where the yellow (cyan) color indicates electron accumulation (depletion).

Insights into the band structures reveal that the temperature drives the redistribution of the electron occupation numbers and the variations of the activated band regions that carry the rapid change information of Berry curvature. Moreover, an increase of Pt thickness will also lead to reconstructions of the electronic band structure and the corresponding Berry curvature distribution in the momentum space, resulting in a reversal of the AHE. These analyses establish that the calculations indeed qualitatively capture the critical physical mechanism of symmetry breaking at the interface, enabling to the significantly enhanced AHE, even though the experimental samples, in terms of thickness and interface roughness, may be more complex than the theoretical model.

To explore the potential topological spin textures at the interface, the samples are examined by the real-space MFM. The MFM signal represents the second derivative of the OOP component of the stray magnetic field, i.e., $d^2H_z/dz^2$. We investigate the evolution of magnetic domains under varying magnetic field at a low temperature (10 K) on a $Cr_5Te_6$ (10 nm)/Pt (3 nm) heterostructure. However, with the magnetic field ranging from -2 T to 0 T, the MFM pattern manifests no obvious magnetic domains, indicating a negligible net magnetic moment (see Figure S7, Supporting Information). It should be noted that the dimensional effect may enhance the antiferromagnetic phase[33] in the $Cr_5Te_6$ layer, leading to a decrease of the stray field. This is further corroborated by the SQUID measurements, indicating that the negligible net magnetic moment is insufficient to induce the notable MFM tip vibrations and generate obvious magnetic domain signals. However, the monotonous image suggests a uniform morphological surface topography, eliminating the contributions of the topographical variances to the MFM pattern. Therefore, the thicker $Cr_5Te_6$ layer (70 nm) with the grown Pt (3 nm) layer is selected for MFM imaging due to its more pronounced net magnetic moment.

Subsequently, the MFM scanning under different magnetic fields is implemented at 10 K to investigate the evolution of the magnetic structure and the emergence of topological spin textures in real space (**Figure 5**a-f). For the OOP magnetization, the MFM detects the areas both inside individual magnetic domains and at the domain walls[55]. The sample exhibits multi-magnetic domains state at 0 T (Figure 5a). When the applied magnetic field reaches 0.5 T, the negative magnetic domains split into island-like patterns (Figure 5b). Further increase in the magnetic field (0.53 T in Figure 5c) makes the island-like domains split into the smaller worm-like domains, accompanied by the appearance of round-shaped bright skyrmions. The worm-like domains are unstable under high fields[56] and almost completely decomposed under 0.6 T (Figure 5d). However, the structures with topologically stable skyrmions do not vanish because it is forbidden from unwinding completely into the plane.[57] The high skyrmions density under 0.6-0.65 T (Figure 5d, e) critically corresponds to the peak of topological Hall effect at about 0.6 T (Figure 5i) which typically arises from the Berry phase acquired by conduction electrons passing through skyrmions. It is expected that further increase in the magnetic field reduce the density of skyrmions (Figure 5f). Subsequently, we present the cross-sectional line profiles of several representative skyrmions. In Figure 5g, h, the plotted data correspond to the white dashed lines in the Figure 5c, d, respectively, recording the variation of the phase modulation in real space. The full-width-at-half-maximum dimension of each skyrmion is approximately 250-300 nm, similar to the size observed by Lorentz TEM in $Cr_{1.53}Te_2$.[25] This remarkable uniformity in both size and magnitude further serves as a robust indicator of the skyrmions. The magnetic domains persist regardless of the magnetic field (marked by black circles in Figure S8, Supporting Information), likely results from the antiferromagnetic phase in $Cr_5Te_6$. To better quantify the relationship between the skyrmions and the MFM results as well as the transport behavior, the topological spin textures phase diagram is show in Figure S9 and Supplementary Note 1 (Supporting Information).

Although the MFM measurement shows no significant magnetic signal in the $Cr_5Te_6$ (10 nm)/Pt (3 nm) heterostructure, the large AHE is still observed in the Hall measurements. Based on the SQUID results and previous analysis,[28] a transition from in-plane anisotropy to nearly magnetic isotropy occurs as the thickness of $Cr_5Te_6$ films decreases. This phenomenon is induced by the weakening of the Coulomb screening in the two-dimensional thin film, resulting in the deflection of the anisotropy axis towards the OOP direction.[26] We depict the Cr spin configuration schematic of a 2 × 1 × 1 magnetic supercell of the $Cr_5Te_6$ under zero field (see Figure S10, Supporting Information), which exhibits a frustrated-like spin structure without the obvious magnetic anisotropy. The degeneration of anisotropy could facilitate the formation of topological spin textures at the interface within the 10-nm-thick $Cr_5Te_6$ heterostructures.[21] Dipole-dipole interactions and exchange interactions, which arise from thickness variations, are additional factors that influence the formation of topological spin textures.

Next, we carry out the atomic simulations utilizing an effective Heisenberg spin system accompany with the interfacial DMI, magnetic anisotropy, and the magnetic anisotropy energy.[58-60] The detailed information of the simulation, including structural and magnetic parameters are shown in the Experimental Section and Supplementary Table S1. The space inversion symmetry is broken by Pt layer at the heterointerface, enabling the interfacial DMI. As a result, a topological spin textures can form at 200 K with a zero magnetic field (Figure 5j). The numerical simulation calculations on the AHE in heterostructures are performed to bridge the Berry phase in real space and Berry curvature in momentum space (Figure 5k). Initially, we postulate the magnetic moment vector of a topological spin texture to be $\boldsymbol{m}(x, y, z)$, with a topological charge as $q_{\text{topo}} = \boldsymbol{m} \cdot (\partial_x \boldsymbol{m} \times \partial_y \boldsymbol{m})$.[61] Thus, the movement of an electron is analogous to the adiabatic evolution along a trajectory with a varying spin quantization axis. Supposing that the local spins are fixed, we have the Schrodinger equation:

$$i\partial_t \psi = \left[\frac{1}{2m}(\hat{\boldsymbol{p}} - q_e \boldsymbol{A})^2 - J\sigma_z\right]\psi, \quad \boldsymbol{\nabla} \times \boldsymbol{A} = \boldsymbol{m} \cdot (\partial_x \boldsymbol{m} \times \partial_y \boldsymbol{m}) \qquad (4)$$

where $q = \pm 1/2$ represents two kinds of spin, $\boldsymbol{A} = (-i/q)U^\dagger \boldsymbol{\nabla} U$ is the Berry connection emerging from the varying local spin, and its curvature exactly corresponding to the topological charge. The emergent magnetic field denoted by $\boldsymbol{A}$ certainly induces a kind of Hall effect, i.e., AHE here (for details see Supplementary

Note 2). While SOC is a well-known source of Berry curvature in momentum space, we emphasize that the topological spin textures also play a crucial role. Numerous studies have shown that in systems such as antiferromagnets, the evolution of the electronic structure interacting with complex spin textures generate significant Berry phase in momentum space that contributes to the AHE.[7, 18, 20] In a recent work on van der Waals magnets ($Fe_{5-x}GeTe_2$), the authors claimed that the topological spin textures (including merons and skyrmions) affect the Berry phase of the charge carriers in both real and reciprocal spaces, that induces the AHE.[62] In our work, through MFM analysis and first-principles calculations, it is demonstrated for the first time that the topological spin textures at the heterointerface interact with the complex electronic band structure, generating an enhanced non-zero Berry curvature, thereby giving rise to the large AHE.

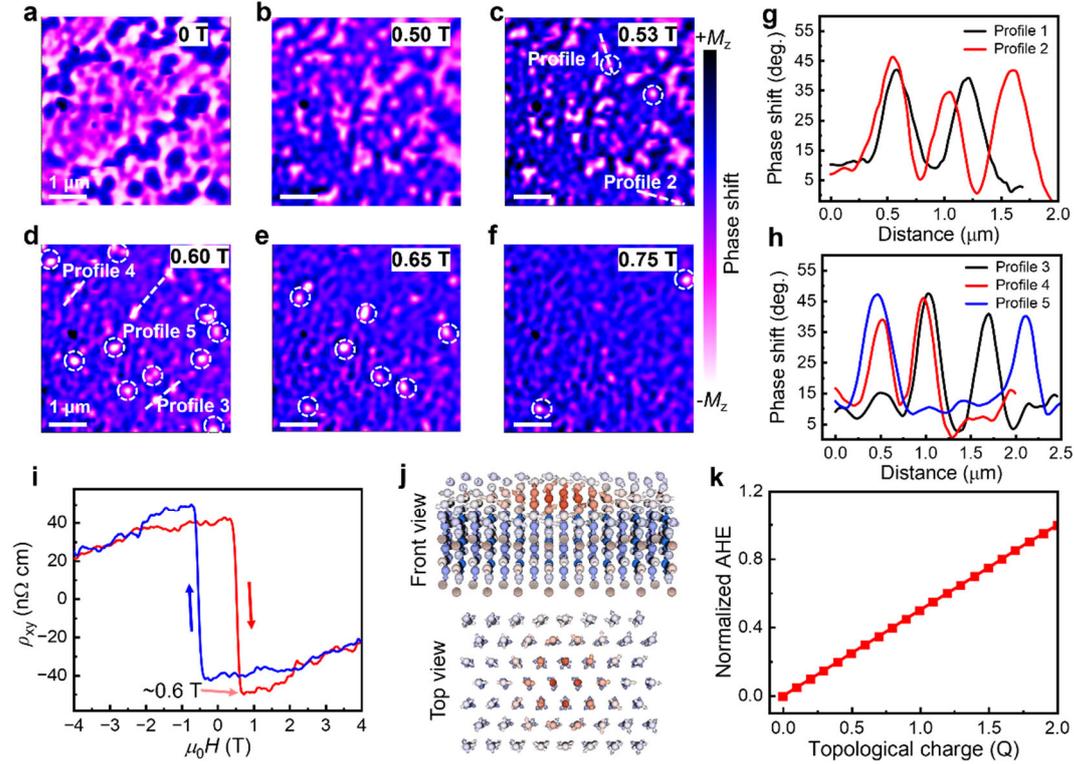

**Figure 5.** Topological spin textures at the interface of $Cr_5Te_6$ (70 nm)/Pt (3 nm) visualized by MFM. a-f) MFM images of the magnetic structures obtained under a field reversing process (field applied OOP) at 10 K. The scanning area is 5 × 5 μm². Dark and bright colors represent positive and negative magnetizations, respectively. As the magnetic field changes, the topologically stable skyrmions (white circle) do not disappear, and the highest density occurs between 0.6-0.65 T (d,e). g,h) Cross-sectional line profiles of representative topological spin textures which are obtained from (c) and (d), respectively. i) The field-dependent Hall resistivity. The hump in the curve originates from the appearance of skyrmions. j) Simulated spin configuration of the topological spin textures generated at the interface of $Cr_5Te_6$ (10 nm)/Pt (3 nm). The diagrams above and below correspond to the front view and top view respectively. The arrows of different colors represent different orientations of spin. k) The relationship between the topological charge and normalized AHE of $Cr_5Te_6$ (10 nm)/Pt (3 nm) through numerical calculations.

## 3. Conclusion

In summary, we have achieved the large anomalous Hall resistivity (114 n$\Omega$ cm at 5 K) in a $Cr_5Te_6$/Pt noncoplanar magnetic heterointerface, which is the highest value reported so far in all the MI/HM heterostructures. Both thermal variations and electron-hole interdiffusion lead to the reconstruction of the Berry curvature, which is verified by our first-principles calculations. The topological spin textures at the interface visualized by the MFM with the support of atomic simulations, highlight the crucial role of these textures in accounting for the AHE. This work opens up a new avenue for exploring the large AHE in heterointerfaces and accelerates the potential applications of MI/HM heterostructures in chiral spintronic devices.

## 4. Experimental Section

*Fabrication of $Cr_5Te_6$/Pt Heterostructures:* All heterostructures were grown on 5 × 5 mm$^2$ $Al_2O_3$ (0001) substrates by the PLD technique. Prior to growth, the substrates were cleaned with acetone, alcohol and deionized water successively and then underwent an annealing process in air at 1100°C for 2 hours. The growth of the heterostructures was performed under a background vacuum of 1 × 10$^{-7}$ mbar. The $Cr_5Te_6$ films were deposited onto the substrates at 500 °C with a deposition rate of 1 nm min$^{-1}$, utilizing a 248-nm KrF excimer laser beam with an average fluence of 1 J cm$^{-2}$ and a repetition rate of 2 Hz.[28] Afterwards, the $Cr_5Te_6$ films were cooled to room temperature to reduce the elemental diffusion in the heterointerface. Subsequently, the Pt films were in-situ grown on the $Cr_5Te_6$ films by the PLD technique. The deposition rate is 0.16 nm min$^{-1}$ (an average fluence of 1.4 J cm$^{-2}$ and a repetition rate of 2 Hz).

*Structural and Magnetic Characterization:* The crystallography was examined by $\theta$-$2\theta$ XRD with Cu K$_\alpha$ radiation on a Rigaku D/MAX-Ultima III instrument. The surface morphology was characterized using an AFM system (NT-MDT). For cross-sectional STEM observation, a sample was prepared using the conventional lift-out method on a Carl Zeiss crossbeam 550L focused ion beam (FIB) combined with scanning electron microscopy (SEM) dual-beam system. The STEM image and the EDX elemental mapping of the $Cr_5Te_6$ (10 nm)/Pt (3 nm) was characterized by an aberration-corrected STEM (FEI Themis Z).

Magnetic properties of 10-nm-thick $Cr_5Te_6$ films were investigated by SQUID. The temperature-dependent magnetization curves were measured in the temperature range of 5 to 300 K with a magnetic field of 100 Oe applied both IP and OOP of the substrates, respectively.

*Transport Measurements:* Relevant transport measurements were conducted by the standard six-probe configuration (Figure 1d). A Cryogen Free Measurement System (CFMS, Cryogenic) was employed with a temperature range from 5 to 300 K. The substrate to the electrode of the device was connected through a gold wire. A constant current of 100 μA and a maximum 4 T magnetic field in the vertical direction were applied during the measurements. In order to eliminate the offset of Hall resistivity caused by contact misalignment, the following formula was used to symmetrize the original data:

$$\rho_{xy} = -t[V_H(+H \to -H) - V_H(-H \to +H)]/2I \qquad (5)$$

*Magnetic Force Microscopic Characterization:* MFM measurements were implemented utilizing a variable temperature system from Attocube equipped with a superconducting magnet (attoDRY2100). All measurements were performed in a vacuum environment, employing a phase modulation methodology in non-contact AC mode. These involved the utilization of a cantilever characterized by a spring constant $k\sim2.8$ N/m and resonant frequency $f\sim75$ kHz. In the AC mode, the cantilever was mechanically excited at its natural resonance frequency with an initial phase. The attractive/repulsive magnetic interaction between the cantilever and sample surface resulted in a negative/positive phase shift. Consequently, the resulting phase mapping imagery provided detailed local insights into the *z*-axis magnetization properties of the magnetic phase.

*First-Principles Calculations:* First-principles simulations were employed to obtain the band structures and the Berry curvature. All the band structures were achieved using the (Vienna Ab-Initio Simulation Package) VASP code within the framework of density functional theory (DFT) and the Perdew-Burke-Ernerhof (PBE) projector augmented was adopted.[63] A vacuum of 20 Å was added to avoid the interaction of adjacent slab calculations. The plane wave basis energy cutoff and the force convergence standard were set to 500 eV and 0.01 eV Å$^{-1}$, respectively. The self-consistent convergence criterion was set to $10^{-4}$ eV. Here, a $2 \times 8 \times 1$ Monkhorst-Pack *k*-points grid was sampled for the structure, which was modeled by a one-atomic-layer of Pt on a $3\times1\times1$ supercell of monolayer $Cr_5Te_6$. SOC was included in all calculations. In addition, DFT wave functions were transformed to maximally localized Wannier functions using the WANNIER90 package.[64] The anomalous Hall conductivity was obtained by[1]

$$\sigma_{AHE} = -\frac{e^2}{\hbar} \int \frac{d^2k}{(2\pi)^2} \Omega_n(\boldsymbol{k}) \qquad (6)$$

where $\Omega_n(\boldsymbol{k})$ is the Berry curvature for the *n*-th subband and the wave vector $\boldsymbol{k}$. Note that $\Omega_n(\boldsymbol{k})$ was along the OOP direction for 2D systems here, and was calculated by[11]

$$\Omega_n(\boldsymbol{k}) = i\Sigma_{n'\neq n} \frac{\langle n'\boldsymbol{k}|\frac{\partial H}{\partial k_x}|n\boldsymbol{k}\rangle \langle n\boldsymbol{k}|\frac{\partial H}{\partial k_y}|n'\boldsymbol{k}\rangle - \langle n'\boldsymbol{k}|\frac{\partial H}{\partial k_y}|n\boldsymbol{k}\rangle \langle n\boldsymbol{k}|\frac{\partial H}{\partial k_x}|n'\boldsymbol{k}\rangle}{[\varepsilon_{n'}(\boldsymbol{k})-\varepsilon_n(\boldsymbol{k})]^2} \qquad (7)$$

where $H$ denotes the Hamiltonian, $\boldsymbol{k} = (k_x, k_y)$ indicates the wave vector, and $|n\boldsymbol{k}\rangle$, $\varepsilon_n(\boldsymbol{k})$ indicate the Bloch wave function and the eigenvalue of the *n*-th subband.

*Atomic Simulations:* In this study, the Cr₅Te₆ served as the fundamental component for our theoretical investigations. The primary interactions accounted for the model included the Heisenberg exchange interaction, the DMI, and the magnetic anisotropy energy. The Hamiltonian expressions for these interactions were strategically formulated as:

$$E = -\sum_i K_i \mathbf{S}_i^2 - \sum_{i \neq j} \left[ J_{ij}^{\text{ex}} \mathbf{S}_i \cdot \mathbf{S}_j + \mathbf{D}_{ij} \cdot (\mathbf{S}_i \times \mathbf{S}_j) \right] \tag{8}$$

where $i$ and $j$ represent the index of atoms, $K_i$ is the anisotropy constant, $J_{ij}^{\text{ex}}$ is the the exchange constant, $\mathbf{D}_{ij}$ is DMI constant vector. For computational approach, thermodynamic properties were approximated through Monte Carlo simulations, ensuring a robust analysis of the magnetic behaviors. The computational tasks were executed using the Vampire software, which facilitated the precise simulation of magnetic materials and their interactions.

**Acknowledgements**

A.S., J.Z., Y.C. and Z.Z. contributed equally to this work. This work was supported by the National Key Research and Development Program of China (Grant No. 2022YFA1402404), the National Natural Science Foundation of China (Grant Nos. T2394473, 624B2070, 62274085, 62122008, 92161201, 12025404, 62374088, and 12074193) and the Fundamental Research Funds for the Central Universities (Grant No. 021014380225).

**Supporting Information**

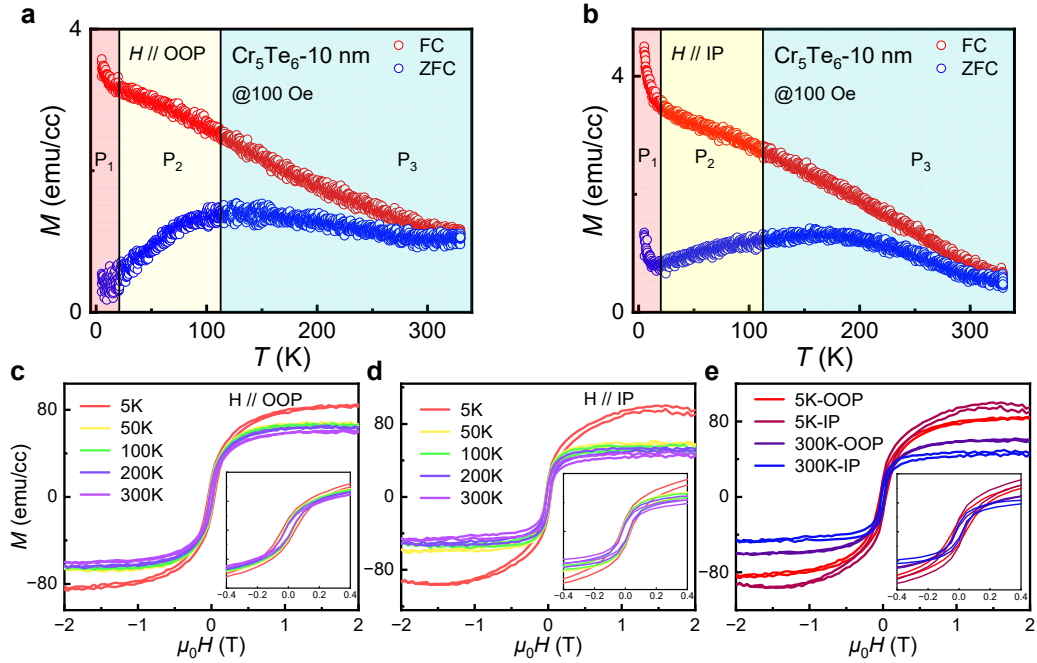

**Figure S1.** Magnetic characterization of the 10-nm-thick $Cr_5Te_6$ thin films. a, b) *M-T* curves of the $Cr_5Te_6$ films with the OOP (a) and IP (b) field, respectively, in which the presence of nodes suggests the occurrence of temperature-dependent magnetic transition processes. c-e) The *M-H* curves of the $Cr_5Te_6$ films with the OOP (c) and IP (d) field at different temperatures, respectively. It demonstrates no obvious magnetic anisotropy (e).

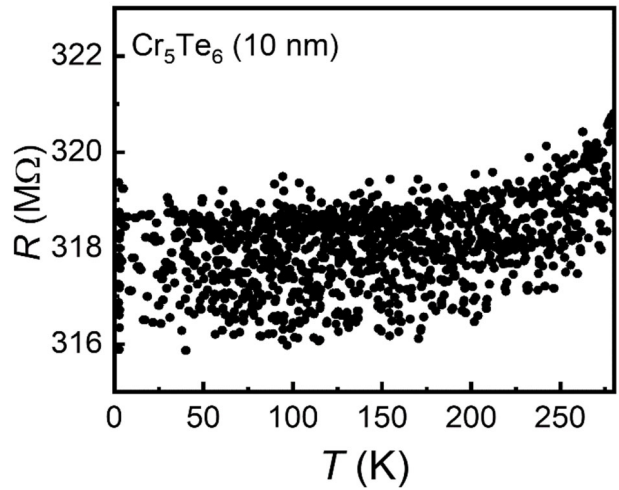

**Figure S2.** Temperature-dependent longitudinal resistance curve of 10-nm-thick $Cr_5Te_6$ films.

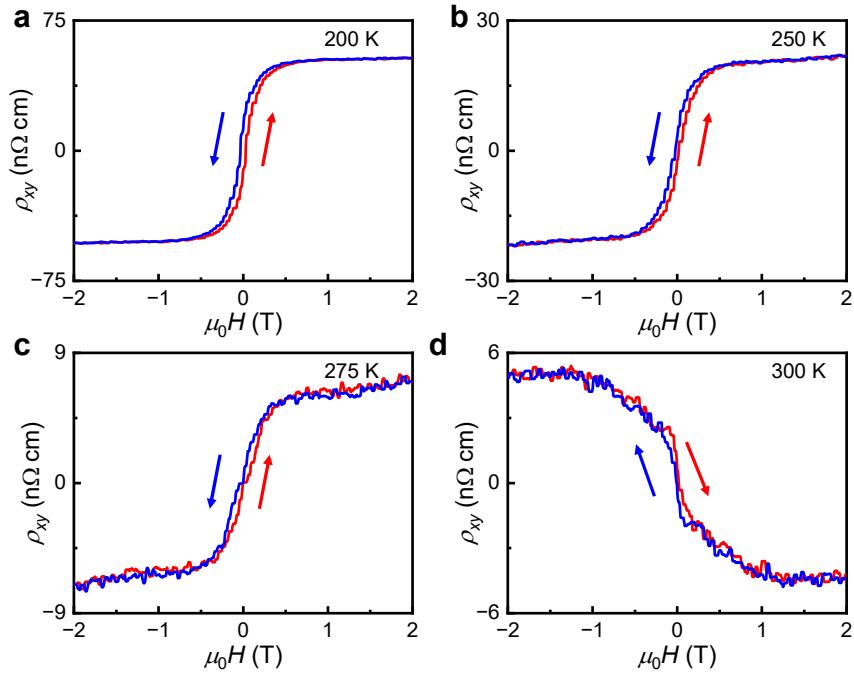

**Figure S3.** The AHE signal reversal of $Cr_5Te_6$ (10 nm)/Pt (3 nm) with varying temperature of (a) 200 K, (b) 250 K, (c) 275 K, and (d) 300 K.

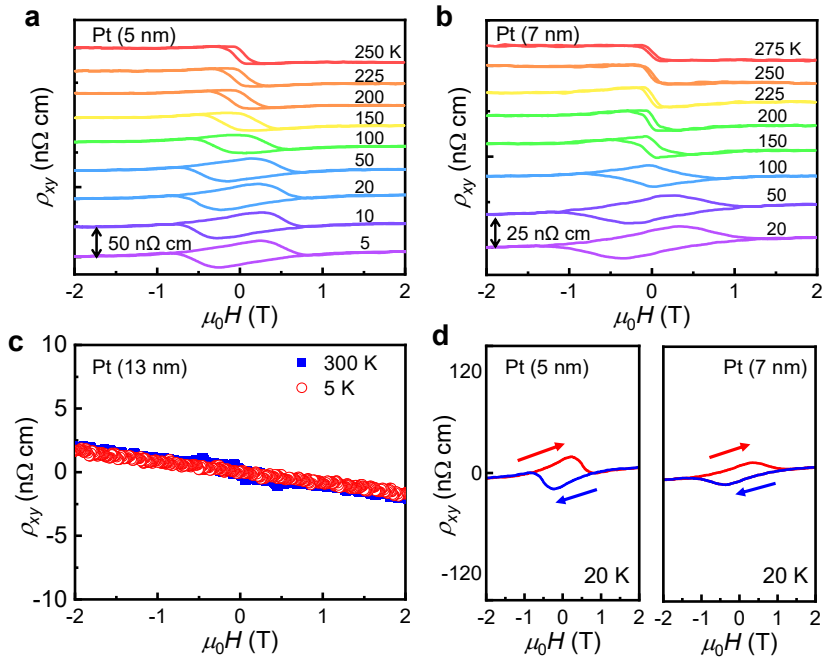

**Figure S4.** Field-dependent Hall resistivity of the $Cr_5Te_6$ (10 nm)/Pt (5 nm) (a), $Cr_5Te_6$ (10 nm)/Pt (7 nm) (b) and $Cr_5Te_6$ (10 nm)/Pt (13 nm) (c) heterostructures, respectively. d) The AHE signal reversal as compared to $Cr_5Te_6$ (10 nm)/Pt (3 nm) heterostructure.

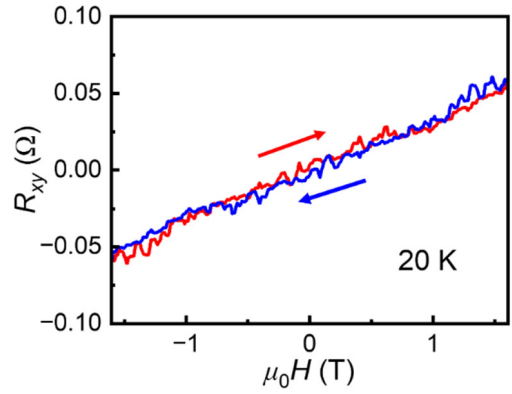

**Figure S5.** The field-dependent Hall resistance of the $Cr_5Te_6$ (10 nm)/Pt (2 nm) heterostructure.

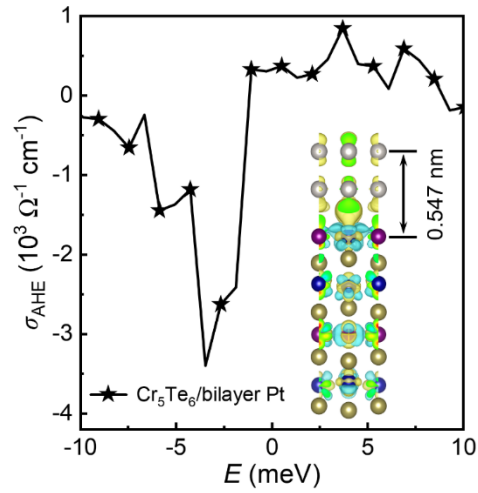

**Figure S6.** Dependence of the anomalous Hall conductivity on the Fermi level in $Cr_5Te_6$/bilayer Pt. The inset shows the charge density difference, where the yellow (cyan) color indicates electron accumulation (depletion).

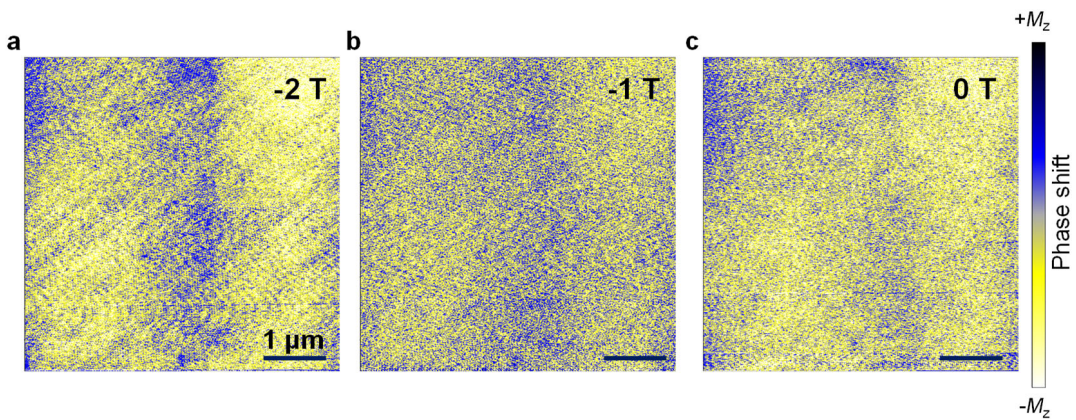

**Figure S7.** MFM characterization of $Cr_5Te_6$ (10 nm)/Pt (3 nm) heterostructure under field sweep processing (field applied OOP) at 10 K. The scanning area is 5 × 5 μm². Dark and bright colors correspond to positive and negative magnetizations, respectively. The weak magnetization results in no significant change in morphology with the applied magnetic field.

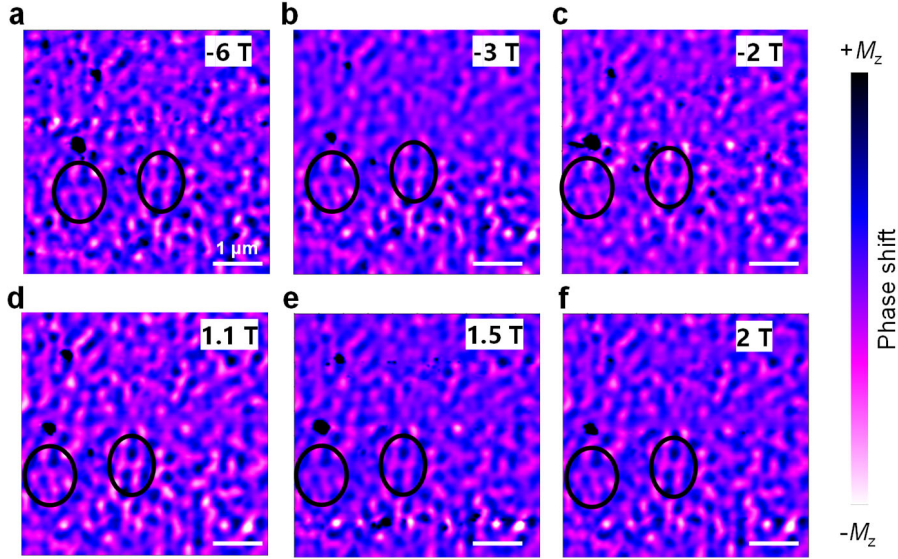

**Figure S8.** MFM images of $Cr_5Te_6$ (70 nm)/Pt (3 nm) heterostructure obtained under the different field at 10 K. The scanning area is $5 \times 5$ μm$^2$. The black circles depicted in the figure represent regions at the interface that remain unaffected under the magnetic field.

**Supplementary Note 1. The Topological Spin Textures Phase Diagram**

The spin state varying with magnetic fields and temperature has been proposed in previous work.[1] This model first classifies the system's ground state into two kinds: stripe domains and uniform orientation, both of which have a vanishing topological charge density. Their energies are compared for varying parameters and the ground state turns out to be the one with the lowest energy. Then, the topological excitation is classified according to the ground state. For these two ground states, it is meron pairs and isolated skyrmions, respectively. Finally, the activation energies of the topological excitation are calculated. For finite temperature, he thermal excitation of these topological solitons must be considered to reproduce the temperature dependence of physical quantities, including the topological charge density and the Hall effect it raises. Note that for a negative excitation energy, the system is in the skyrmion crystal phase. Finally, the phase diagram is plotted in Figure S9a, and its profile at T = 50 K and 100 K is exhibited in Figure S9b. The Hall effect is proportional to the topological charge density.

$$\rho_{\text{Hall}} = \frac{P\hbar q}{ne^2} \tag{1}$$

where $\rho_{Hall}$ is the Hall resistivity, $P$ is the spin polarization of the current, $n$ is the carrier concentration, and $e$ is the elementary charge.

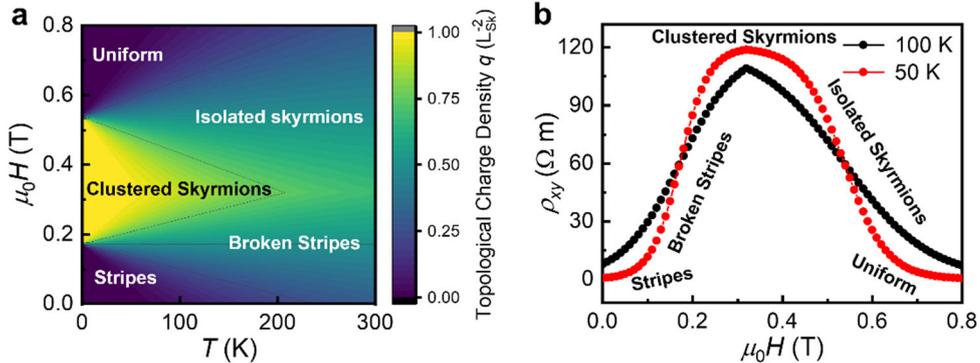

**Figure S9.** a) The topological spin textures phase diagram of $Cr_5Te_6$/Pt heterostructures. b) The corresponding profile at 50 and 100 K, respectively.

**Table S1.** Structural and magnetic parameters of atomic simulations

| Parameters | Value | Unit |
|---|---|---|
| Type I Cr atomic-spin-moment | 2.24 | $\mu_B$ |
| Type II Cr atomic-spin-moment | 4.24 | $\mu_B$ |
| Te atomic-spin-moment | 0 | |
| Cr-Cr exchange constant | 8~30 (estimated) | meV |
| Cr-Cr 2$^{nd}$ exchange constant | -5~-3 (estimated) | meV |
| Anisotropy constant | 4~5×10$^5$ (estimated) | J m$^{-3}$ |
| Lattice vector-a | 6.913 | Å |
| Lattice vector-b | 3.970 | Å |
| Lattice vector-c | 12.44 | Å |
| DMI constant | 0.2~2 | meV |

**Supplementary Note 2. AHE Enabled by Topological Spin Textures**

To assess the impact of the magnetic topological structure of $Cr_5Te_6$ on the performance of devices denoted by AHE, we develop a simplistic physical model. Initially, we can define a topological structure as a region encompassing a magnetic skyrmion, where we postulate the magnetic moment vector field to be $\bm{m}(x, y, z)$, leading to the presence of a topological number expressed as,

$$q_{\text{topo}} = \bm{m} \cdot (\partial_x \bm{m} \times \partial_y \bm{m}) \qquad (2)$$

As electrons traverse through this topological structure, the migration velocity is influenced due to the action of the Berry curvature.[2-3] Assuming the electrons' initial velocity to be $v_{\text{ori}}$, and the original momentum to be $\hat{\bm{p}}$, the relationship can be formulated as,

$$v_{\text{shif}} = v_{\text{Berry}} + v_{\text{ori}}, \ \hat{\bm{p}} \to \hat{\bm{p}} - e\bm{A} \qquad (3)$$

where $v_{\text{shif}}$ denotes the velocity after adjustment. The deviation in electron velocity is attributable to the magnetic topological structure originates from the Berry curvature. To precisely calculate this deviation, it is necessary to compute the Berry curvature.

In a modern language, the vortex is termed as Berry curvature and the vector potential as Berry connection. This language gives an insightful view on the effect of magnetic vector potential when it appears in the momentum operator, which is associated with phase change of waves[2]. When an electron moves in a magnetic film, the local spin interacts with the electron spin and induces the precession of the electron spin. The Schrodinger equation is

$$i\partial_t\psi = \left(\frac{1}{2m}\hat{p}^2 - J\hat{\boldsymbol{\sigma}}\cdot\boldsymbol{m}\right)\psi \tag{4}$$

where $\psi$ is the electron's wave function, $\hat{\boldsymbol{\sigma}}$ is the electron spin, and $\boldsymbol{m}$ is the local spin.

The Berry connection will be exposed after a unitary transformation is applied to the equation to simplify the spin operator. The transformation is done by aligning the spin quantization axis at each point with the local spin there. The movement of an electron is thus analogous to the adiabatic evolution along a trajectory with a varying spin quantization axis. Supposed the local spins are fixed, we have the Schrodinger equation[3]:

$$i\partial_t\psi = \left[\frac{1}{2m}(\hat{\boldsymbol{p}} - q_e\boldsymbol{A})^2 - J\sigma_z\right]\psi, \quad \boldsymbol{\nabla}\times\boldsymbol{A} = \boldsymbol{m}\cdot(\partial_x\boldsymbol{m}\times\partial_y\boldsymbol{m}) \tag{5}$$

where $q = \pm 1/2$ denotes two kinds of spin, $\boldsymbol{A} = (-i/q)U^\dagger\boldsymbol{\nabla}U$ is the Berry connection emerging from the varying local spin, and its curvature exactly corresponds to the topological charge.

$U$ is the unitary transformation that rotates $\boldsymbol{m}$ to the z-axis:

$$\widehat{U}(\boldsymbol{r},t) = \exp\left\{-i\frac{\Theta(\boldsymbol{r},t)}{2}\hat{\boldsymbol{\sigma}}\cdot[\hat{\boldsymbol{z}}\times\boldsymbol{m}(\boldsymbol{r},t)]\right\} \tag{6}$$

By the unitary transformation or a gauge transformation which can more accurately capture the local spin textures and their influence on the Berry curvature, topological invariants, and subsequent transport effects. The emergent magnetic field denoted by $\boldsymbol{A}$ surely causes a kind of Hall effect, i.e., AHE in our case.

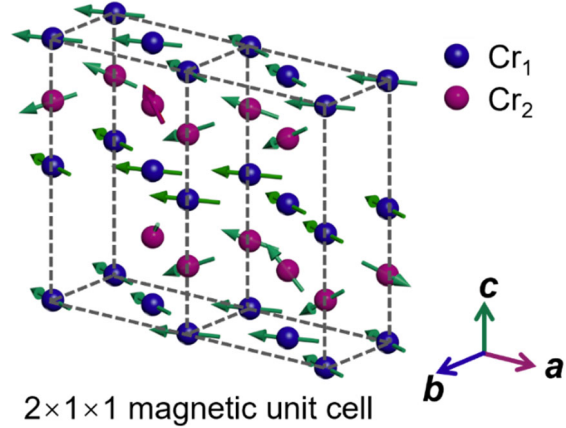

**Figure S10.** Schematic diagram of the spin configuration of the Cr for $2\times 1\times 1$ $Cr_5Te_6$ magnetic supercell without magnetic field, which exhibits frustrated-like spin structures without the significant magnetic anisotropy.